\documentclass[fleqn,usenatbib]{mnras}

\newcommand{\tsim}{\ensuremath{\sim}}
\newcommand{\nifs}{\ensuremath{^{56}\rm{Ni}}}

\newcommand{\Msun}{M$_{\odot}$}

\newcommand{\kms}{km\hspace{0.25em}s$^{-1}$}
\newcommand{\ergs}{erg\hspace{0.25em}s$^{-1}$}

\newcommand{\OI}{\mbox{O\hspace{0.25em}{\sc i}}}

\newcommand{\NaI}{\mbox{Na\hspace{0.25em}{\sc i}}}

\newcommand{\MgI}{\mbox{Mg\hspace{0.25em}{\sc i}}}

\newcommand{\CaII}{\mbox{Ca\hspace{0.25em}{\sc ii}}}

\newcommand{\FeII}{\mbox{Fe\hspace{0.25em}{\sc ii}}}

\newcommand{\Nifs}{$^{56}$Ni}

\newcommand{\Mej}{$M_{\textrm{ej}}$}

\newcommand{\KE}{\ensuremath{E_{\rm K}}}

\newcommand{\eg}{e.g.\ }
\def\gsim{\mathrel{\rlap{\lower 4pt \hbox{\hskip 1pt $\sim$}}\raise 1pt \hbox {$>$}}}
\def\lsim{\mathrel{\rlap{\lower 4pt \hbox{\hskip 1pt $\sim$}}\raise 1pt \hbox {$<$}}}
\usepackage[T1]{fontenc}
\usepackage{ae,aecompl}
\usepackage{graphicx}
\usepackage{epstopdf}
\usepackage{amsmath}
\usepackage{amssymb}
\usepackage{enumerate}

\title[Modelling the Type Ic SN 2004aw]{Modelling the Type Ic SN 2004aw:
a Moderately Energetic Explosion of a Massive C+O Star without a GRB}
\author[P. A. Mazzali, et al.]{P. A. Mazzali,$^{1,2}$
\thanks{E-mail: P.Mazzali@ljmu.ac.uk}, 
D. N. Sauer$^{3}$, 
E. Pian$^{4,5}$, 
J. Deng$^{6}$, 
S. Prentice$^{1}$,
S. Ben Ami$^{7}$  
\newauthor S. Taubenberger$^{2,8}$ and 
K. Nomoto$^{9}$
\\
\\
$^{1}$Astrophysics Research Institute, Liverpool John Moores University, IC2, 134 Brownlow Hill, Liverpool L3 5RF, United Kingdom \\
$^{2}$Max-Planck-Institut f\"ur Astrophysik, Karl-Schwarzschild-Str. 1, 85748 Garching bei M\"{u}nchen, Germany\\
$^{3}$German Aerospace Center (DLR), Institute of Atmospheric Physics, 82234
Oberpfaffenhofen, Germany \\
$^{4}$INAF IASF-Bo, via Gobetti 101, 40129 Bologna, Italy\\
$^{5}$Scuola Normale Superiore, Piazza dei Cavalieri, 7, 56126 Pisa, Italy \\
$^{6}$National Astronomical Observatories, CAS, 20A Datun Road, Chaoyang District, Beijing 100012, China\\
$^{7}$Smithsonian Astrophysical Observatory, 60 Garden St., Cambridge MA-02138, USA \\
$^{8}$European Southern Observatory, Karl-Schwarzschild-Str. 2, 85748 Garching bei M\"{u}nchen, Germany\\
$^{9}$IPMU, Kashiwa, 277-8583, Japan}
\date{Accepted ... Received ...; in original form ...}
\pubyear{2015}

\begin{document}
\label{firstpage}
\pagerange{\pageref{firstpage}--\pageref{lastpage}}
\maketitle

\begin{abstract}
An analysis of the Type Ic supernova (SN) 2004aw is performed by means of models
of the photospheric and nebular spectra and of the bolometric light curve.
SN\,2004aw is shown not to be ``broad-lined'', contrary to previous claims, but
rather a ``fast-lined'' SN\,Ic. The spectral resemblance to the narrow-lined
Type Ic SN\,1994I, combined with the strong nebular [\OI] emission and the broad
light curve, point to a moderately energetic explosion of a massive C+O star.
The ejected $^{56}$Ni mass is $\approx 0.20$\,\Msun. The ejecta mass as
constrained by the models is $\sim 3-5$\,\Msun, while the kinetic energy is
estimated as \KE\,$\sim 3-6 \times 10^{51}$ ergs. The ratio \KE/\Mej, the
specific energy which influences the shape of the spectrum, is therefore
$\approx 1$.  The corresponding zero-age main-sequence mass of the progenitor
star may have been $\sim 23-28$\,\Msun. Tests show that a flatter outer density
structure may have caused a broad-lined spectrum at epochs before those observed
without affecting the later epochs when data are available, implying that our
estimate of \KE\ is a lower limit. SN\,2004aw may have been powered by either a
collapsar or a magnetar, both of which have been proposed for gamma-ray
burst-supernovae. Evidence for this is seen in the innermost layers, which
appear to be highly aspherical as suggested by the nebular line profiles. 
However, any engine was not extremely powerful, as the outer ejecta are
more consistent with a spherical explosion and no gamma-ray burst was detected
in coincidence with SN\,2004aw.
\end{abstract}


\section{Introduction}
\label{sec:intro}

Type Ic supernovae (SNe) are H-/He-poor core-collapse supernovae
\cite{clow97,fil97,Matheson2001,Modjaz2016}. Significant diversity can be found
among the He-poor SNe whose data have been published
\citep{Bianco2014,Taddia2015,Lyman2016,Prentice2016}. For example, models of the
gamma-ray burst (GRB) SN\,1998bw \citep{iwa98} indicate that it ejected
\Mej\,$\sim 10$\,\Msun\ of material with kinetic energy \KE\,$\sim 4\times
10^{52}$ erg, and synthesised $\sim 0.4$\,\Msun\ of \Nifs, which powered the
light curve. These results suggested a progenitor of $\sim 40$\,\Msun. In
contrast, SN\,1994I \citep{ric96} was less luminous and energetic. Radiation
transport models \citep{sau06} show that it ejected only $\sim 1$\,\Msun\ of
material with \KE\,$\sim 10^{51}$\,erg. Both the \KE\ and the mass of \Nifs\
synthesised by SN\,1994I ($M_\mathrm{Ni}\sim0.08$\,\Msun) are similar to
ordinary SNe\,IIP, which suggests a progenitor mass of $\sim 15$\,\Msun. A star
of this mass must undergo severe envelope stripping to result in anything other
than a SN\,IIP.  \citet{hachinger12} showed that small amounts of He and H are
sufficient to transform a spectrum from H-/He-poor to H-poor/He-rich and from
H-poor/He-rich to H-/He-rich, respectivel. How this mass loss occurs is not
fully understood, but strong binary interaction is considered to be the most
likely way to strip a progenitor of its outer layers
\citep{nom95,Eldridge2013}. 

Given the large range of \Mej\ of SNe\,Ic we can infer that a wide range of
progenitor stars can end their lives as SNe\,Ic, depending on their evolution.
Many of the best observed SNe\,Ic are luminous, massive, and energetic, and 
belong to an extreme subclass sometimes called hypernovae. These are
characterised by early spectra showing very broad absorption features indicative
of very high ejecta velocities \citep{maz02,mazzali10}.  Some of these SNe have
been discovered in coincidence with long-duration gamma-ray bursts (GRBs)
\citep[e.g.,][]{gal98,maz06a,Ashall2017}. While the collapsar scenario,
involving the formation of a black hole and the subsequent accretion on to it of
additional material from the stellar core \citep{mac99,woo06,fry07} has enjoyed
much success, the fact that all GRB-SNe have \KE\,$\sim 10^{52}$\,erg, and that
the collimation-corrected energy of the associated GRBs is always much smaller
than the SN \KE\ has been suggested to indicate that GRB-SNe could be energised
by magnetars \citep{mazzali14,Greiner2015,Ashall2017}.  SN\,2006aj, which was
associated with an X-ray flash, was modelled as the explosion of a star of
$M_{\rm ZAMS}\sim 20$ M$_{\sun}$, and a magnetar was claimed to have energised
the explosion also in this case \citep{maz06b}. Magnetars may therefore be
responsible for most, perhaps all SNe associated with relativistic events, which
would be in keeping with the original suggestion that the collapsar is a failed
SN \citep{mac99}. We also know SNe which have a high-velocity, high-energy tail
to their ejecta, which causes broad absorption lines, but were not associated
with a GRB \citep[\eg SN\,1997ef,][]{maz00b}. In some cases the mass and
energies involved may have been too small \citep[\eg SN\,2002ap,][]{maz02}; in
other cases the orientation of the event may have been unfavourable for the
detection of a possible GRB \citep[\eg SN\,2003jd,][]{val08,maz05}. 

Extracting SN properties requires good observational data as well as explicit
modelling of these data (spectra and light curves), which implies some amount of
work and has been done for only a very small number of SNe.  Although
significant uncertainties affect the estimates of mass and energy obtained even
with this method, \citet{mazzali13} show that properties derived using scaling
arguments can easily lead to much larger uncertainty, while properties derived
using simplistic approaches are not to be trusted at all. It is therefore
important not only to extract the properties of the most extreme SNe, but to map
the entire range of \Mej, \KE, and \Nifs\ mass in order to understand how
SNe\,Ic are produced and what is the relation to other types of
stripped-envelope SNe, which may show different distributions of properties
\citep[\eg][]{Prentice2017a}. 

\begin{figure}
  \includegraphics[width=88mm]{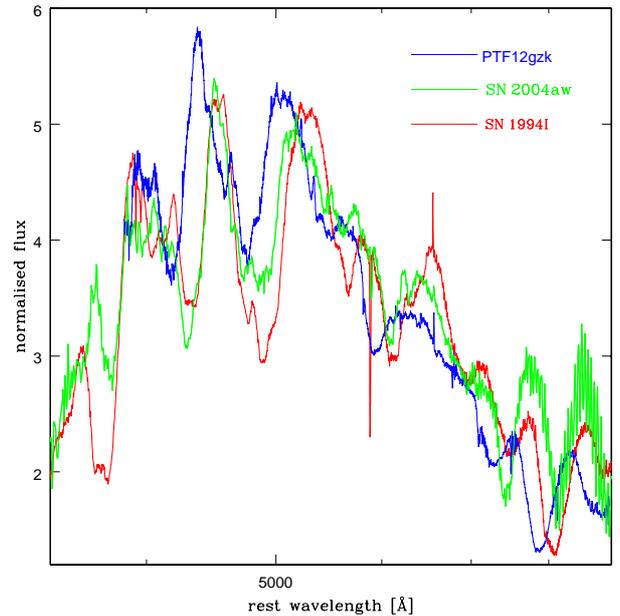}
  \caption{Spectra of SN\,2004aw (2004 Mar 24, 1 day after $B$-band maximum), 
  1994I (1994 Apr 4, 4 days before $B$-band maximum) and 
  PTF12gzk (2012 Aug 2, 2 days before $B$-band maximum), showing the close 
  similarity in spectral shape and line width and the shift in line velocity.
  PTF12gzk has the highest velocities, and SN\,1994I has the lowest, especially 
  if the early epoch of the spectrum shown here is taken into account.}
  \label{fig:cf94I04aw}
\end{figure}

SN\,2004aw is an interesting SN to study because it appears to be an
intermediate case and it was very well observed \citep{tau06}. Under the
classification scheme presented in \citet{Prentice2017a} it is classified as a
SN\,Ic-6. It was spectroscopically similar to SN\,1994I and PTF12gzk
\citep{benami2012}, which was characterised by high line velocities ($>20,000$
\kms). SN\,2004aw is intermediate between these two SNe in velocity. In Figure
\ref{fig:cf94I04aw} we compare near-maximum spectra of the three SNe. Despite
the large velocity difference between the spectra, it is apparent that none of
them show the extreme line blending that is typical of broad-lined SNe\,Ic. The
light curve of SN\,2004aw is somewhat broader than that of SN\,1994I, and 
depending on the reddening that is assumed it may be also be more luminous.  
SN\,2004aw may therefore represent an intermediate type of event, and as such it
provides a good opportunity to further our understanding of He-poor SNe.  

In this paper we describe our one-dimensional (1-D) models for the photospheric
and nebular spectra of SN\,2004aw, and for the bolometric light curve. Using our
1-D radiative transfer codes, we constrain the ejected mass, kinetic energy, and
ejected $^{56}$Ni mass of SN\,2004aw, which are then compared with those of
other Type Ic SNe. 

We analysed SN\,2004aw in the spirit of abundance tomography \citep{mazzali14}.
In Section 2 we describe our models for the early-time spectra, and discuss how
they were used to establish a density distribution. We then proceed to the
nebular spectra, and show how these can be used to determine the mass and
density of the ejecta at low velocities, as well as their composition, in
particular with respect to the amount and distribution of \Nifs. We also show
how the profile of the emission lines can be used to optimise the model (Section
3). We then use the model of the ejecta obtained through spectral modelling,
including the distribution of \Nifs, to compute a synthetic bolometric light
curve, which we compare to the bolometric light curve of SN\,2004aw constructed
from the available photometry (Section 4).  Finally, in Section 5 we discuss our
results, and place SN\,2004aw in the context of type Ic SNe.

Following \citet{tau06}, a distance modulus of $\mu=34.17$ for the host galaxy,
NGC 3997, was used. A combined Galactic plus local reddening of $E(B-V)=0.37$ is
assumed throughout this paper.

\section[]{Models for the Photospheric Spectra}
\label{sec:early}

We modelled a series of photospheric spectra of SN\,2004aw using our Montecarlo
SN spectrum synthesis code. Spectra with epoch between 1 and 28 days after $B$ 
maximum were selected from \citet{tau06}.

Our method is based on a Monte-Carlo solution of the line transfer that was
developed by \citet{abb85}, \citet{maz93}, \citet{luc99}, and \citet{maz00a}. 
The models assume an inner boundary (``photosphere'') of velocity $v_{\rm
ph}(t)$ for an epoch $t$ relative to the explosion where all radiation is
emitted as a blackbody continuum of temperature $T_{\rm BB}$. This is generally
a good assumption for early epochs, and typically works until epochs of $3-4$
weeks after maximum for oxygen-dominated SN\,Ic ejecta. The ejecta are described
by a density distribution (an explosion model) and by a composition, which can
vary with depth. The gas in the ejecta and the radiation field is assumed to be
in radiative equilibrium. Photon energy packets are followed through the ejecta,
where they undergo scattering events.  For line scattering, a branching scheme
based on transition probabilities in the Sobolev-approximation allows photons to
be emitted in transitions different from those in which they were absorbed. In
order to account for the energy of photons that are scattered back into the
photosphere (line blanketing) $T_{\rm BB}$ is adjusted in an iterative procedure
to obtain the desired emergent luminosity, $L$. The emergent spectrum is
obtained from a formal solution of the transfer equation using the source
functions that are derived from the Monte-Carlo simulation. The code has been
used for a number of SNe\,Ic \citep[\eg][]{mazzali13}. 

\begin{figure}
  \centering
  \includegraphics[width=86mm]{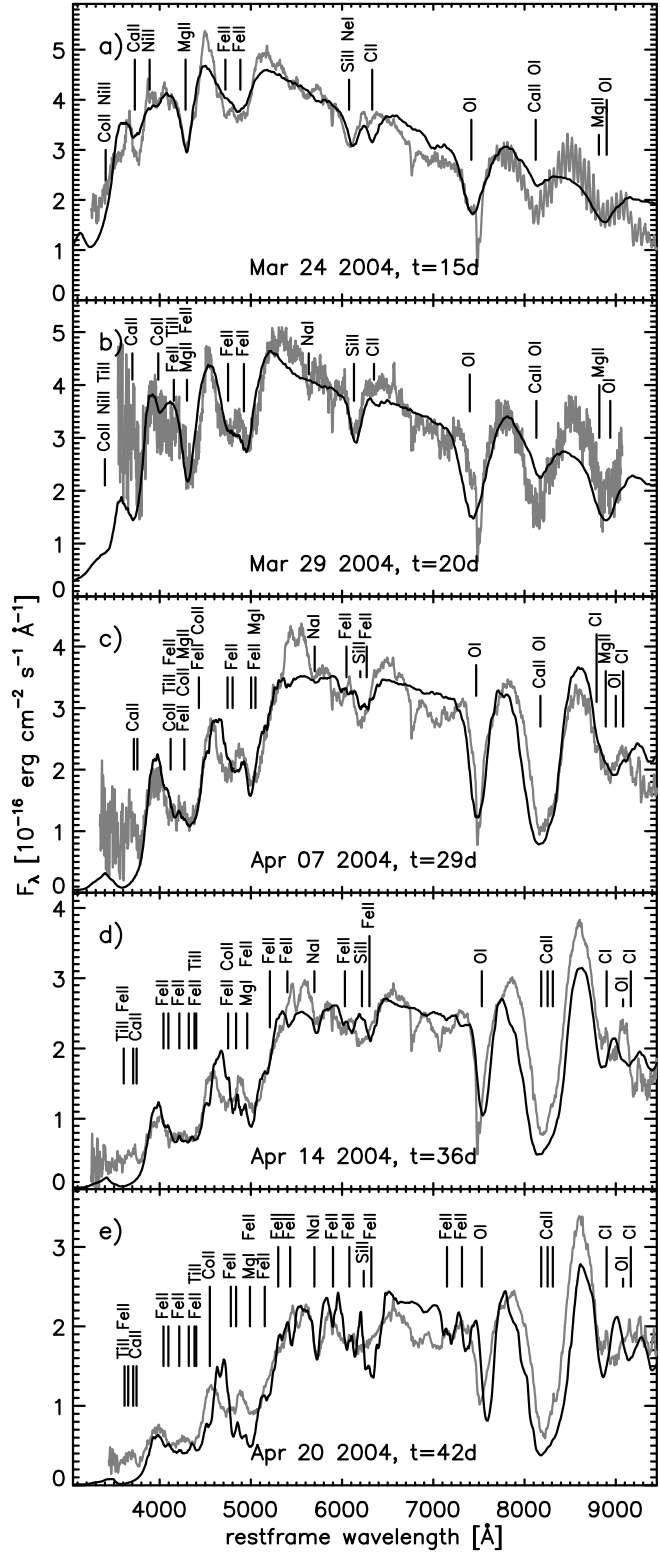}
  \caption{Model spectra (black lines) compared to the respective observed 
  spectra of SN\,2004aw \citep{tau06} at different epochs (gray lines). 
  The line labels refer to the most prominent ions causing the respective 
  absorption features.
  The epochs refer to the time after the assumed explosion date.
  }\label{fig:spectra}
\end{figure}

Early-time spectra probe the outer part of the ejecta, and can be used to
establish both the abundance and the density distribution. In the case of
SN\,2004aw the density structure above $5000\,${\kms} was modelled based on the
1-D hydrodynamical model CO21, which was developed for SN\,1994I \citep{iwa94}.
The density was scaled up by a factor of $3$ in order to achieve enough optical
depth in the lines. This approach is justified by the overall spectral
similarity between SN\,2004aw and SN\,1994I at similar photospheric epochs
\ref{fig:cf94I04aw}and by the good match of our model spectra to the observed
ones (see Figure \ref{fig:spectra}). This simplistic approach is necessary
because of the lack of a grid of explosion models at different masses and
energies. On the other hand, rescaling is justified because CO cores of massive
stars tend to have self-similar density profiles even if they differ in mass (G.
Meynet, priv. comm.). This rescaling can lead to some additional uncertainty, so
we are very generous with our error estimates below. Homologous expansion is
assumed at all times, so the ejecta can be conveniently described using velocity
as a comoving coordinate.

As the supernova ejecta expand the line-forming region of the supernova recedes
to progressively lower velocities.  Subsequent epochs are therefore modelled by
adding shells below the previous inner boundary while retaining the density and
composition of the layers described by models of earlier epochs, following the
approach called ``abundance tomography'' \citep{ste05}.

Our models require a time from explosion so that the density structure can be
appropriately rescaled. SN\,2004aw was discovered late, only 5 days before $B$
maximum, so a direct determination of the epoch of explosion is not possible. 
\citet{tau06} showed that the light curves of SN\,2004aw are broader than those
of SN\,2002ap, whose risetime was determined through modelling to be $\sim 10$
days \citep{maz02}. SN\,2004aw reached $V$ maximum $\sim 3$ days after $B$
maximum \citep{tau06}. Bolometric maximum was intermediate between these two
(see Section \ref{sec:lc}).  We adopted for SN\,2004aw a rise-time to $B$
maximum of 14 days. Epochs used in the models are shown in Figure
\ref{fig:spectra} and listed in Table \ref{tab:para} along with other modelling
parameters. We did not correct for time-dilation caused by the small redshift
$(cz = 4906\,$\kms) since the uncertainty in the epoch (at least $\pm 1$ day) is
larger than the correction. 

Figure~\ref{fig:spectra} shows our best-fitting model spectra ({\em black
lines}) compared to the observed spectra ({\em gray lines}). This ejecta model
has \tsim $3$ M$_{\sun}$ of material above 5000 \kms. We also tested ejecta
models with different mass in an effort to break the parameter degeneracy that
affects the light curve modelling. We found that the SN\,2004aw spectra can be
reproduced reasonably well only if the density scaling factor of CO21 is between
$\sim$ 2 and $\sim$ 4.  Models with larger mass have excessively high line
velocities which cause the absorption features to appear too broad. In
particular, the oxygen line at $\sim7400\,${\AA} becomes significantly too broad
if too much material is present at high velocities. This line is highly
saturated and therefore fairly insensitive to the O abundance within reasonable
limits, meaning that the strength and shape of the absorption feature is
primarily set by the density structure. The lower limit for the mass above any
photosphere is constrained by the amount of material needed to generate notable
P-Cygni features.

In the following paragraphs we discuss the properties of the individual models
in more detail. The primary input parameters for all early-time spectral models
are summarized in Table~\ref{tab:para}. 

\begin{table}
  \caption{Model parameters for the early-time models}
  \label{tab:para}
  \centering
  \begin{tabular}{ccccc}
    Date 	& $t$ &   $L$   	 & $v_{\rm ph}$ & $T_{\rm BB}$  \\
         	& [d] & [erg/s] 	 &    [km/s]    &      [K]      \\
    \hline
    24 Mar 2004 & 15 & $4.04\times 10^{42}$ & 11100 & $8.9 \times10^{3}$ \\
    29 Mar 2004 & 20 & $3.65\times 10^{42}$ &  9600 & $8.0 \times10^{3}$ \\
    07 Apr 2004 & 29 & $2.97\times 10^{42}$ &  8100 & $6.6 \times10^{3}$ \\
    14 Apr 2004 & 36 & $2.30\times 10^{42}$ &  6600 & $5.9 \times10^{3}$ \\
    21 Apr 2004 & 43 & $1.91\times 10^{42}$ &  5000 & $6.0 \times10^{3}$ \\
    \hline
  \end{tabular}
\end{table}

The first spectrum that we modelled is from March 24, 2004, $1\,$day after
$B$-band maximum and $t=15\,$d after explosion.  Fig.~\ref{fig:spectra}a shows a
comparison of the model spectrum to the observed one. The model with
$L=4.04\times10^{42}\,${\ergs} and $v_{\rm ph}=11\,100\,${\kms} reproduces most
of the observed features although the re-emission features are not strong enough
in some places. All \CaII\ absorptions, the H\&K feature at \tsim$3750\,${\AA}
and the IR triplet at \tsim$8100\,${\AA}, are somewhat weaker in the model than
in the observation. To fit these features, however, a much higher Ca abundance
would be required which cannot be accommodated by the later spectra. Therefore
it is likely that the model at this epoch overestimates the ionization,
resulting in too much \ion{Ca}{iii} at the expense of \ion{Ca}{ii}. The
composition used to model this spectrum includes equal parts of O and C
(\tsim\,$45$ per cent by mass), $7$ per cent Ne, a total of $1.3$ per cent
intermediate-mass elements (Mg, Si, S, Ar) including $4\times10^{-4}$ per cent
Ca. Additionally, we use $0.062$ per cent Fe-group elements consisting of $0.05$
per cent {\nifs}, $0.012$ per cent ``stable'' Fe (i.e., Fe not produced via the
{\nifs} decay chain) and traces of Ti and Cr.

Figure~\ref{fig:spectra}b shows the spectrum on March 29, 2004, compared to the
model spectrum. The epoch, $6\,$days after maximum light, corresponds to
$t=20\,$d after the assumed explosion date. The model requires a luminosity of
$L= 3.65\times10^{42}\,$\ergs. The pseudo-photosphere is located at $v_{\rm
ph}=9600\,${\kms} with a temperature of $T_{\rm BB}=8060\,$K for the underlying
blackbody. The composition is similar to the previous model although less stable
Fe is needed to match the Fe features because more {\nifs} has decayed to Fe.

The next epoch we modeled is 2004 April 7, 15 days after maximum light and
$t=29\,$d after explosion (Fig.~\ref{fig:spectra}c). The luminosity used in this
model is $L=2.97\times10^{42}\,$\ergs\ at a photospheric velocity of $v_{\rm
ph}=8100\,${\kms}. The resulting temperature of the photosphere is $T_{\rm
BB}=6640$ K. The shell near the photosphere contains still \tsim$84$ per cent C
and O but somewhat more intermediate-mass elements, and $1.1$ per cent {\nifs}. 
This spectrum appears to be much redder than the earlier ones. In addition to a
lower temperature, blocking by a large number of iron group lines suppresses the
UV and blue flux in this and later spectra. The fit to the observed spectrum is
acceptable although the double-peaked re-emission features between $5400$ and
$6700\,${\AA} are too weak in the model. At this and the subsequent epochs the
models also fail to reproduce the absorption features around $6800$ and
$7200\,${\AA}. Based on the models it cannot be uniquely asserted if this
discrepancy is due to missing elements in the composition or if the ionization
balance is incorrectly determined.  The assumption of the thermal photosphere
absorption additionally overestimates the flux in this wavelength region.

The following spectrum (Fig.~\ref{fig:spectra}d) was taken on 2004 April 14, 22
days after maximum light and $t=36\,$d after explosion.  The luminosity used
here is $L=2.30\times10^{42}\,$\ergs. The photospheric velocity is $v_{\rm
ph}=6600\,${\kms}, which leads to a blackbody temperature of $T_{\rm
BB}=5950\,$K.  The composition used for this fit is still very similar to the
previous epoch, with a slightly higher {\nifs} mass fraction (1.6 per cent). 
The \OI\ feature at $7400\,${\AA} appears stronger in the model than in the
observation. Unfortunately, the observed spectra show an atmospheric absorption
feature at $7500\,${\AA} which happens to coincide with the \OI\ feature.  We
decided to use the observed spectrum in which this feature has not been removed
because it makes the uncertainty in the shape of the \OI\ absorption more
apparent. 

The last epoch we modelled with the photospheric method
(Fig.~\ref{fig:spectra}e) is 20 April 2004, 28 days after maximum light and
$t=43\,$d after explosion.  The model requires a change in the abundance
pattern.  C and O are reduced to $10$ and $35$ per cent, respectively, Si is
increased to $15$ per cent and we require $18$ per cent of {\nifs}. The inner
boundary of this model is located at $v_{\rm ph}=5000\,${\kms}. The overall
shape of the spectrum is well reproduced although the velocity of some
absorptions is underestimated by the model, indicating that the change in
composition may actually occur at somewhat higher velocities than assumed here.
The uncertainty in the shape of the density structure in this transition region,
however, makes it difficult to find a better match. To improve the fit we used
an intermediate shell at $5800\,${\kms} which allows us to smoothen the
transition somewhat, although the lack of an observed spectrum between Apr 14
and 21 leaves this shell relatively unconstrained.

\section[]{Models for the nebular spectra}
\label{sec:neb}

More information about the properties of the deeper layers of the ejecta can be
obtained from nebular-epoch spectra. At late phases, when the optical depth of
the ejecta has dropped to below 1, the innermost ejecta produce emission lines,
whose strengths and profiles can shed light on the core-collapse process.

The nebular spectra of Type Ic SNe are usually dominated by a strong
[\ion{O}{i}] $\lambda\lambda$6300, 6364 line, and SN\,2004aw is no exception. 
In the case of the GRB/SN\,1998bw, the nebular spectra were used to infer the 
aspherical distribution of different elements in the ejecta and hence the
aspherical nature of the explosion. The narrow [\ion{O}{i}]
$\lambda\lambda$6300, 6364 line and the contrasting broad [\ion{Fe}{ii}]
emission suggested a polar-type explosion viewed near the axis
\citep{maz01,mae02}. For SN\,2003jd, which was luminous but did not show a GRB,
the double-peaked profile of the [\ion{O}{i}] line witnesses the disc-like
distribution of oxygen in an explosion similar to that of SN\,1998bw but viewed
closer to the equatorial plane \citep{maz05}.

In contrast, other SNe\,Ic show no sign of asphericity in their late-time
profiles.  Examples include a low-energy SN such as SN\,1994I \citep{sau06} and
a narrow-lined SN with higher energy such as SN\,2006aj \citep{mazzali07a}.

\begin{figure}
 \centering 
 \includegraphics[width=\columnwidth]{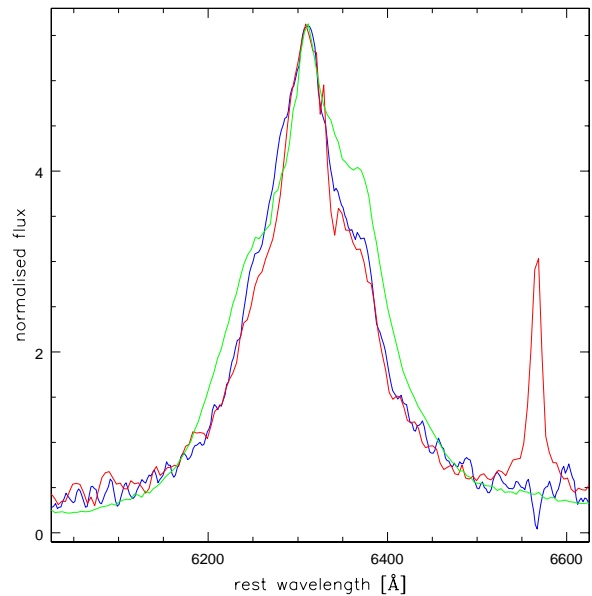}
 \caption{The [\ion{O}{i}] $\lambda\lambda$6300, 6364 emission line in the 
  spectrum of SN\,2004aw, $235\,$d after explosion (blue line) compared to the 
  same line in the spectrum of the GRB/SN\,1998bw obtained on 21 May 1999, 388
  rest-frame days after the explosion (red) and that of SN\,2002ap obtained on
  16 Sept 2002, 229 days after explosion (green). }
\label{fig:cf04aw98bw02ap} 
\end{figure}

The [\ion{O}{i}] $\lambda\lambda$6300, 6364 emission profile in SN\,20004aw is
remarkably similar to that of SN\,1998bw (Figure \ref{fig:cf04aw98bw02ap}),
indicating that SN\,2004aw must have been significantly aspherical. A detailed
comparison shows that SN\,2004aw has a broader emission base, and the narrow
core that characterised SN\,1998bw emerges only at low velocities. This suggests
a similar morphology for the two SNe, but more extreme in SN\,1998bw, and a
similar viewing angle.  Below we attempt to model the nebular spectrum of
SN\,2004aw in order to define its properties.

\begin{figure}
 \centering 
 \includegraphics[width=\columnwidth]{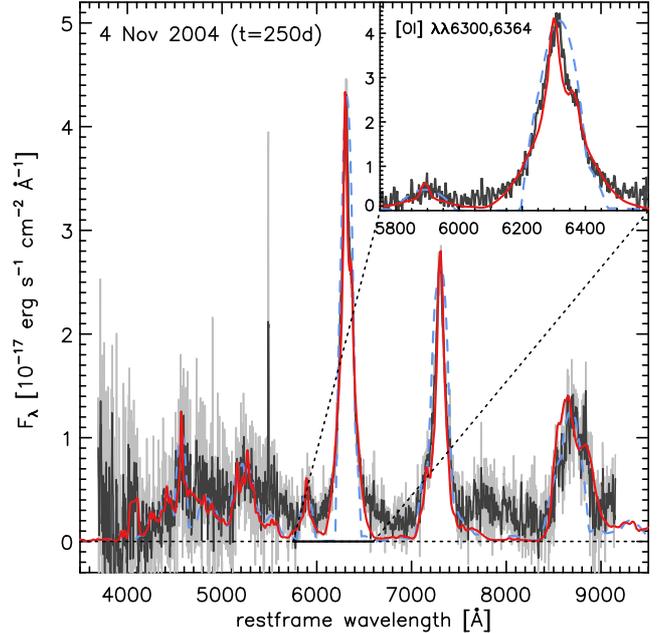}
 \caption{The observed nebular spectrum (light gray line),
  $235\,$d after explosion compared to the model spectra. The one-zone model
  is indicated by the blue dashed line, the solid red line represents the
  improved multi-zone model.  For a better comparison we also show a smoothed
  version of the observed spectrum in dark gray. The inset shows a blow-up of 
  the [\ion{O}{i}] $\lambda\lambda$6300, 6364 emission. }
\label{fig:neb} 
\end{figure}

Only the first nebular spectrum of SN\,2004aw, obtained on 4 Nov 2004, $\sim
235$ days after maximum, has sufficiently high signal-to-noise ratio that
several lines can be detected. The spectrum is shown in Figure \ref{fig:neb}
along with synthetic models. The emission lines that can be identified are, from
blue to red: \ion{Mg}{i}] $\lambda$4571; [\ion{Fe}{ii}] multiplets near
5200\,\AA, the strength of which is related to the peak luminosity of the SN; a
weak \NaI\,D line at 5890\,\AA; the very strong [\ion{O}{i}]
$\lambda\lambda$6300, 6364 emission; the strong line near 7250\,\AA, which
contains both [\ion{Ca}{ii}] $\lambda\lambda$7291, 7320 and [\ion{Fe}{ii}]
lines; a weaker \ion{Ca}{ii} IR triplet near 8600\,\AA, which also contains some
\ion{C}{i}.

We modelled the spectrum using a code \citep{maz01} that computes $\gamma$-ray
and positron deposition following the decay of $^{56}$Ni to $^{56}$Co and hence
to $^{56}$Fe, and then balances the collisional heating induced by the
thermalisation of the fast particles produced by the deposition with cooling via
line emission in non-LTE, following the prescriptions of \citet{axe80}. The code
can be used in two modes, as discussed in \citet{mazzali07a}: a one-zone version
simply assumes a nebula of constant density, with a sharp outer boundary at a
velocity the value of which is selected at input, as are the mass of the nebula 
and the elemental abundances within it. In a more sophisticated version the code
still assumes spherical symmetry but allows for a radial variation of density
and composition, and does not require an outer boundary. In this version the
diffusion of $\gamma$-rays is followed with a Montecarlo method \citep{capp97}.
This version is useful to test the prediction of explosion models as well as to
investigate the radial distribution of mass and abundances in detail when the
observed line profiles are sufficiently well determined. 

In order to determine the global properties of the nebular spectrum of
SN\,2004aw we began by constructing a first, coarse model using the one-zone
approach, adjusting the abundances to fit the shape of the prominent emission
features. The synthetic spectrum we computed is indicated by a blue dashed line
in Figure \ref{fig:neb}, where it is compared to the observed spectrum (light
grey). The darker line represents the observed spectrum that has been smoothed
to emphasise emission features especially in the blue. Assuming a typical rise
time of 14 days, we used an epoch of 250 days after explosion for the
calculation. The outer velocity adopted in the calculation that gives a
reasonable fit to most emission lines is $5000\,$\kms.  This shows that the
nebular spectra originate in a region deeper than the innermost layers studied
by means of early-phase spectroscopy, but the separation between the two regions
is small.  The mass contained within $5000\,$\kms\ in the model is $\sim
1.8$\,\Msun.  The dominant element is oxygen (1.3\,\Msun). The mass of \nifs\
required to reproduce the \FeII\ lines and at the same time to excite all other
transitions is 0.17\,\Msun. This is larger than in the low-mass SN\,Ic 1994I,
and similar to energetic, broad-lined SNe\,Ic such as 1997ef \citep{maz00b} or
2006aj \citep{maz06b}. The carbon mass is small, only 0.2\,\Msun, as determined
by the strength and shape of the \CaII-dominated feature near $8500\,${\AA}. A
small C/O ratio is quite common in SNe\,Ic. The Mg mass is also small,
0.004\,\Msun, despite the relative strength of \MgI] $4571\,$\AA. 

The blue dashed line in the blow-up in Fig.~\ref{fig:neb} shows in detail the
[\OI] line in the one-zone model.  When the observed line profile is viewed in
detail it is clear that it has a composite structure. It shows a broad base,
which can be described by a symmetric emission with limiting velocity
$5000\,$\kms, and a narrow core, of width $\sim 2000$\,\kms, superimposed on
it.  As we showed above, it is similar to the profile observed in SN\,1998bw. A
similar type of profile was also observed in the BL-SN\,Ic 2002ap, and it may be
interpreted as a signature of asphericity \citep{mazzali07b}. As discussed in
that paper, the profile can be produced by the superposition of a narrow
emission core and either a broad, flat-topped profile, which can originate in a
shell-like distribution of oxygen, or a double-peaked profile such as what is
expected from an equatorial distribution of oxygen viewed on or close to the
equatorial plane \citep{maz05}. The shell+core configuration would be globally
spherically symmetric, while the disc+core one would be aspherical. It is not
possible to distinguish between these two scenarios based only on the profile of
the [\OI] line. In the case of SN\,1998bw, the simultaneous observation of broad
[\FeII] emission lines favoured the disc+core solution. In the case of
SN\,2004aw, the Fe lines are not sufficiently well observed for us to be able to
determine which scenario is favoured. Still, the similarity of the [\OI] profile
suggests that the inner parts of SN\,2004aw behaved like SN\,1998bw. The [\OI]
emission profile is SN\,2002ap was 
also similar. Interestingly, the [\OI] line in SN\,2002ap was the broadest of
the three SNe.  The broad emission base suggests that the intermediate-velocity
layers ($v \sim 5-10000$\,\kms) were more spherical in SN\,2002ap than in either
SN\,1998bw and 2004aw. The weaker central emission core also indocates that any
central density enhancement was also less extreme.  

An aspherical distribution of matter requires a detailed model of the
explosion.  Here we used our stratified code to test the spherically symmetric
scenario. We used the density and abundance distributions obtained from the
modelling of the early-time spectra at velocities down to $5000\,$\kms, the
velocity of the photosphere of the last of the early-time spectra (20 April
2004). Therefore, we freely adjusted both the density and the abundances below
that velocity, trying to optimise the detailed fit of the line profiles, and in
particular that of [OI] $6300, 6364\,$\AA\ and its narrow core. 

The solid red line in Figure \ref{fig:neb} shows the multi-zone synthetic
spectrum. The narrow core has a characteristic velocity of $\sim 2000\,$\kms,
and there is a clear discontinuity in line emission at velocities between 2000
and $4000\,$\kms. In order to reproduce such a feature, it was necessary to
introduce a discontinuity in the density profile. The density decreases slightly
below $4000\,$\kms, reaching a minimum at $3000\,$\kms, and then it increases
sharply again in deeper layers. This distribution places a core containing
$\approx 0.25$\,\Msun\ of material below $3000\,$\kms. The composition of this
core must be dominated by oxygen if the sharp peak of the [OI] line is to be
reproduced, but it also must contain \nifs\ in order to excite oxygen.  The
one-zone model already suggests that a large fraction of the \nifs\ is located
at $v < 5000\,$\kms. As a result, the \nifs\ distribution is smooth, peaking at
4-5000\,\kms, but the distribution of oxygen is bimodal: the O abundance is
large outside of $6000\,$\kms, and again below $3000\,$\kms. The density
structure used to fit the \ion{O}{i} feature also produces reasonable fits to
the other feature, which however are quite noisy.  The nebular model includes a
total of $0.2 M_{\sun}$ of \nifs. The total ejected mass is $3.9 M_{\sun}$, of
which 0.85\,\Msun\ are located at $v < 5000\,$\kms.

\begin{figure}
  \centering
  \includegraphics[width=\columnwidth]{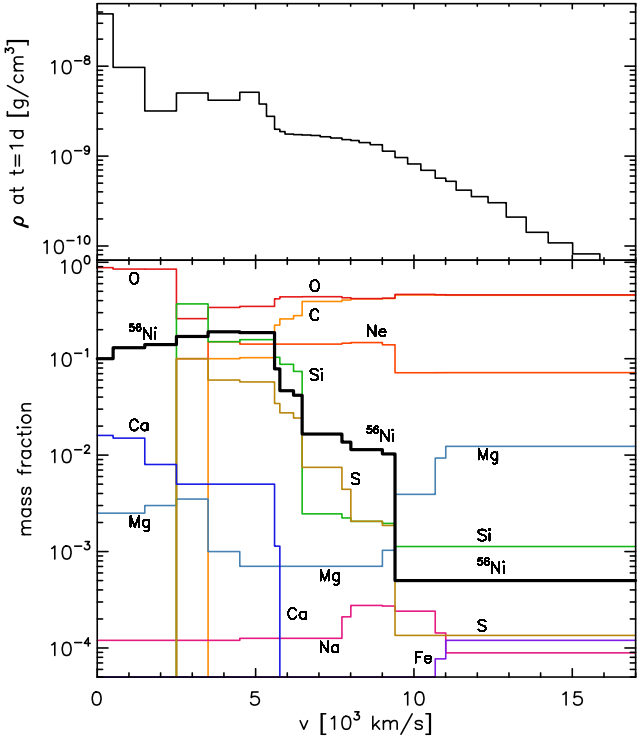}
  \caption{\label{fig:composition} The density (upper panel) and composition
  structure (lower panel) we used to model SN~2004aw. The nebular spectrum
  probes the inner part of the ejecta out to \tsim$7000\,${\kms}. The lowest
  inner boundary for the photospheric spectra is located at $5000\,${\kms}.  }
\label{fig:tomography} 
\end{figure}

The density distribution that we finally derived is shown in Figure
\ref{fig:tomography}. It shows the sharp increase of density at the
lowest velocities, indicating the presence of in inner region dominated by
oxygen. 

Classical, one-dimensional explosion models do not predict the presence of any
material below a minimum velocity, which represents the position of the
"mass-cut": material below this mass cut forms the compact remnant and is not
ejected. Light curve studies had already suggested the presence of an inner
dense core of material in some SNe\,Ic, in particular the massive and energetic
ones linked to gamma-ray bursts \citep{Maeda2003}. The nebular spectra of both
SNe\,2004aw and 1998bw indicate that the inner region is dominated by oxygen, a
constitutent of the stellar core, rather than by products of nucleosynthesis.
The most likely explanation for this distribution of mass and abundances is that
the low-velocity material was ejected in the low-energy part of an aspherical
explosion. In the case of SN\,1998bw the entire [\OI] profile is sharply peaked,
indicating a grossly aspherical explosion, while in the case of SN\,2004aw the
sharp component of the emission is narrower, suggesting that the asphericity
affected only the innermost parts of the ejecta.

\section[]{Bolometric Light Curve}
\label{sec:lc}

\subsection{Construction of the Bolometric Light Curve}

We evaluated a pseudo-bolometric light curve of SN2004aw as follows.  The
magnitudes reported in Table 2 of Taubenberger et al. (2006) were dereddened for
Galactic extinction using $E_{B-V} = 0.37$ and the extinction curve of 
\citet{Cardelli89} and converted to fluxes according to \citet{fukugita95}. 
Note that those magnitudes are {\it all} corrected for the contribution of the
host galaxy, although this was evaluated with different methods for different
telescopes: the KAIT magnitudes were obtained via aperture photometry on
template-subtracted images,  while for all the rest PSF photometry was used
within SNOOPY, which includes a correction for the local background.

The light curves in the $JH$ and $K$ bands cover a much more limited time
interval than the $UBVRI$ ones; therefore, we assumed that their temporal
behaviour at epochs where observations are not available follows that of the
$I$-band. Similarly, we assumed that the $U$- and $B$-band fluxes at the
penultimate and last epochs follow the same general temporal trend as the other
optical bands.

The $UBVRIJHK$ monochromatic light curves were splined with a time resolution of
1 day. The monochromatic fluxes in the various bands were linearly interpolated
and extrapolated redwards and bluewards of the $K$- and $U$-band filter
boundaries, respectively. Broad-band spectral energy distribution were thus
constructed at each epoch and integrated in flux over the range 3000-24000\,\AA.
Any contribution at wavelengths outside the above range was ignored and is
likely not to exceed 5-10\%. The resulting pseudo-bolometric luminosities
corresponding to the epochs of the available photometry are plotted in Figure
\ref{fig:LC}.

In order to estimate the uncertainties, the errors associated with the optical
and NIR photometry were propagated by adding them in quadrature. For the epochs
where only optical photometry was available, the errors on the NIR photometry
were estimated based on the errors at the epochs when data were available.

The resulting light curve is consistent with that reported in \citet{tau06} when
account is taken of the somewhat different treatment of flux at the boundaries
of the wavelength range adopted for integration.

\subsection{Light Curve Model}

We compare the bolometric light curve obtained as discussed above with models
computed with the SN Montecarlo light curve code discussed first in
\citet{capp97} and expanded in \citet{maz00a}. Based on a density structure and
a composition, the code computes the propagation and deposition of gamma-rays
and positrons (using the same description as the nebular spectrum code),
transforms the deposited energy into radiation and follows the propagation of
the optical photons in the expanding ejecta. Although it is based on simple
assumptions about the opacity \citep{maz00a,hoeflich95} it does yield a
reasonable representation of the bolometric light curve
\citep[\eg][]{mazzali13}. 

A synthetic bolometric light curve was computed for the ejecta density and
abundance distribution that was obtained from spectral modelling as outlined
above. Above 5000\,\kms\ the model is a scaled-up version of model CO21, which
gave good fits to SN\,1994I.  The photospheric spectra of SN\,2004aw resemble
those of SN\,1994I at similar epochs (with respect to the $B$-band maximum),
apart from having higher line velocities. This suggests a similar density
structure and a similar mass-to-kinetic energy ratio for both SNe for the part
of ejecta that is responsible for spectral lines in the photospheric phase.
Below 5000\,\kms, however, we adopted the density derived in Sec. \ref{sec:neb}
to reproduce the nebular spectra. The resulting explosion model has an ejected
mass of $\sim 4$\,\Msun, an energy \KE\,$\sim 4 \times 10^{51}$\,erg, and a
\Nifs\ mass of $\approx 0.2$\,\Msun. 

\begin{figure}
 \includegraphics[width=88mm]{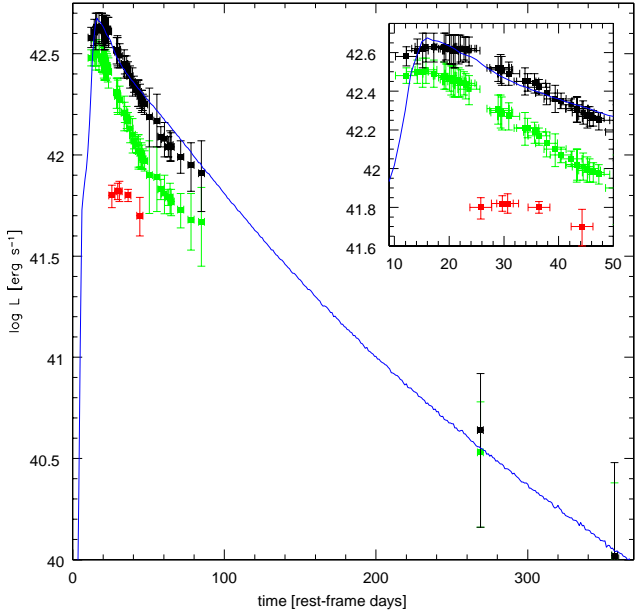}
  \caption{\label{fig:LC}The synthetic light curve computed using the 
  spectroscopic results for SN\,2004aw (blue line) compared to the 
  pseudo-bolometric light curve of SN 2004aw (black squares). The optical data 
  used to construct the pseudo-bolometric light curve are shown as green 
  symbols, while the NIR photometry is shown as red symbols. The inset is a
  blow-up of the peak phase.}
\end{figure}

Figure\,\ref{fig:LC} shows the synthetic bolometric light curve compared to the
observed one constructed as described above.  The final bolometric light curve
uses as reference the time of $B$ maximum, since the time of the explosion is
not accurately known. Bolometric maximum occurs 1-3 days after $B$-band maximum
according to our calculation. A pre-discovery limit was obtained on 2004 Mar 13,
which precedes $B$-band maximum by $\sim$ 10 days. The light curve of SN\,2004aw
is relatively broad for a SN\,Ic. The rise time of the $B$-band light curve is
$\sim 14$ days for the GRB/SN\,1998bw \citep{gal98} and $\sim 9$ days for the
XRF/SN\,2006aj \citep{pia06}, whose explosion dates are known from the
accompanying GRBs. In previous theoretical studies of Type Ic SNe the assumed
(or model-constrained) rise time $\sim 9-11$ days for SN\,1994I
\citep{iwa94,bar99,sau06} and $\sim 9$ for SN\,2002ap \citep{maz02}, and as much
as 20 days for SN\,1997ef \citep{maz00a}. In the case of SN\,2004aw a rise time
of 14 observed days is used, which matches the epoch of bolometric maximum in
the synthetic light curve.

Although the model does not capture all undulations in the observed light curve,
it does reproduce its overall flux level and the time of maximum, indicating
that the model that was used has reasonable values of mass, energy, and \Nifs\
mass. We take this as a confirmation of the spectroscopic results. A possible
reason for the discrepancies is the approximate treatment of the (gray) opacity,
but this was not an issue in other cases \citep[\eg][]{mazzali13}. Another
possible source of uncertainty is that we based our calculations on a rescaled
rather than a real explosion model, and this rescaling was significant. Thus our
model may not fully capture the properties of a significantly more massive
explosion. We therefore use generous error bars on our estimates of mass and
velocity.

We can conservatively estimate that, in order to reproduce the bolometric light
curve of SN\,2004aw, the ejecta mass should be in the range $\sim 3-5$\,\Msun.
For smaller masses we can expect the peak to occur too early, while for larger
masses the opposite would be the case.

\section{Discussion and Conclusions}
\label{sec:disc}

Our models suggest that SN\,2004aw was the explosion of a massive C+O star that
ejected $\sim 4$\,\Msun\ of material. If we consider a range of possible remnant
masses, from a neutron star with mass 1.5\,\Msun\ up to a black hole with mass
3\,\Msun\, this leads to an exploding CO core of $\sim 4.5-8$\,\Msun. This
points to a progenitor star of $M_{\rm ZAMS}\sim 23-30$\,\Msun, in the context
of the pre-supernova evolution models of \citet{nom88} and \citet{has95} (see
also \citealt{woo86}). The estimate of $M_{\rm ZAMS}$ may be modified if the
large mass-loss required to remove the H and He envelopes is taken into
consideration. For massive single stars, this strongly depends on the uncertain,
mass-dependent mass-loss rate in the Wolf-Rayet stage. For example,
\citet{woo93,woo95} found that all their models with $M_{\rm ZAMS}\ga
35$\,\Msun\ evolve to a narrow final C+O core mass range of $\sim 2-4$\,\Msun\,
which may be marginally compatible with SN\,2004aw. In contrast, \citet{pol02}
obtained a final C+O core mass in excess of 10\,\Msun\ for $M_{\rm ZAMS} >
30$\,\Msun\ at solar metallicity, adopting updated mass-loss rates, while models
with $M_{\rm ZAMS} <30$\,\Msun\ retain a substantial He envelope before
explosion. Near-infrared data of SN\,2004aw \citep{tau06} clearly rule out the
presence of significant amounts of He \citep{hachinger12}.  SN\,2004aw may have
been the outcome of common envelope evolution in a massive close binary, a
scenario first proposed by \citet{nom94,nom95} for Type Ic SNe in general and
SN\,1994I in particular. This was developed for low-mass SNe\,Ic but it may also
work at large masses.  Alternatively, binary evolution with stable mass transfer
may lead to a C+O progenitor, but this mechanism may work only at lower masses
than what is required for SN\,2004aw \citep{yoon2015}. At high masses,
Wolf-Rayet wind mass-loss may strip stars of their outer H and He envelopes and
produce the progenitors of massive SNe\,Ic. 

\begin{figure}
 \includegraphics[width=88mm]{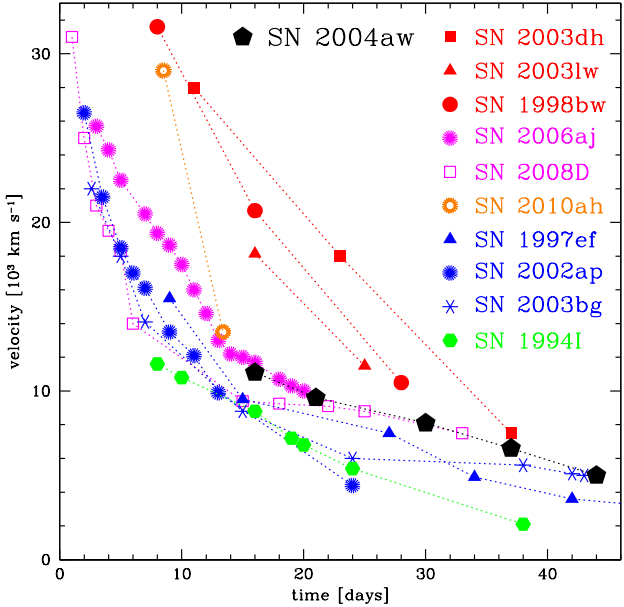}
  \caption{\label{fig:Sivel}Model photospheric velocities in 
  SN\,2004aw (pentagons) and in other SNe\,Ib/c.}
\end{figure}

The explosion of SN\,2004aw was moderately energetic, resulting in a kinetic
energy $\KE \sim 4\times 10^{51}$\,erg. The ratio of kinetic energy and ejected
mass is $\sim 1\,[10^{51}$\,erg/\Msun], similar to low-energy explosions such as
SN\,1994I \citep{sau06} and significantly smaller than that of low-mass
hypernovae such as SN\,2002ap \citep{maz02}. SN\,2004aw is often referred to as 
a broad-lined SN\,Ic, even though its early-time spectra are quite different
from those of hypernovae such as SN\,2002ap \citep{tau06} and very similar to
those of a low-mass SN\,Ic like SN\,1994I. In Figure \ref{fig:Sivel} we examine
the behaviour of the model photospheric velocity as a function of time compared
to other SNe\,Ib/c with similar modelling. We do not use observed line
velocities as those measurements can be highly uncertain because of line
blending and broadening, and because line absorption happens at different
velocities for different lines. SN\,2004aw has no early information but at the
times when it was observed the photospheric velocity follows the behaviour of
SNe like 2006aj or 2008D, which do not have a very large $\KE/$\Mej. SN\,2004aw
also follows the behaviour of SN\,1994I, although it does so at higher
velocities. 

Could the earlier spectra of SN\,2004aw have shown broad lines? We mentioned
above that the presence of broad lines is mostly the result of the density slope
in the outermost layers. We can modify the density slope in the outermost
layers, and test what the spectrum would have looked like at an epoch preceding
the observed ones, making sure that the modification does not affect the
earliest available spectrum. 

\begin{figure}
 \includegraphics[width=88mm]{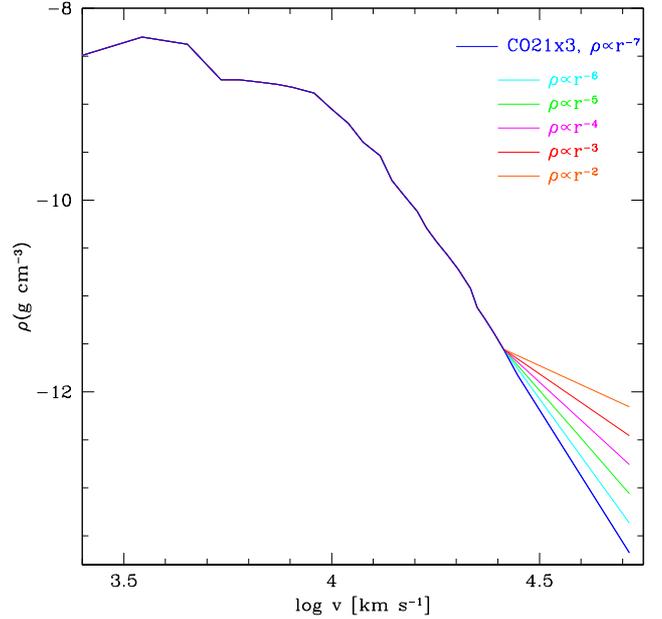}
  \caption{\label{fig:densityslope}Modified density profiles used to test the 
  earliest properties of SN\,2004aw.}
\end{figure}

\begin{figure}
 \includegraphics[width=88mm]{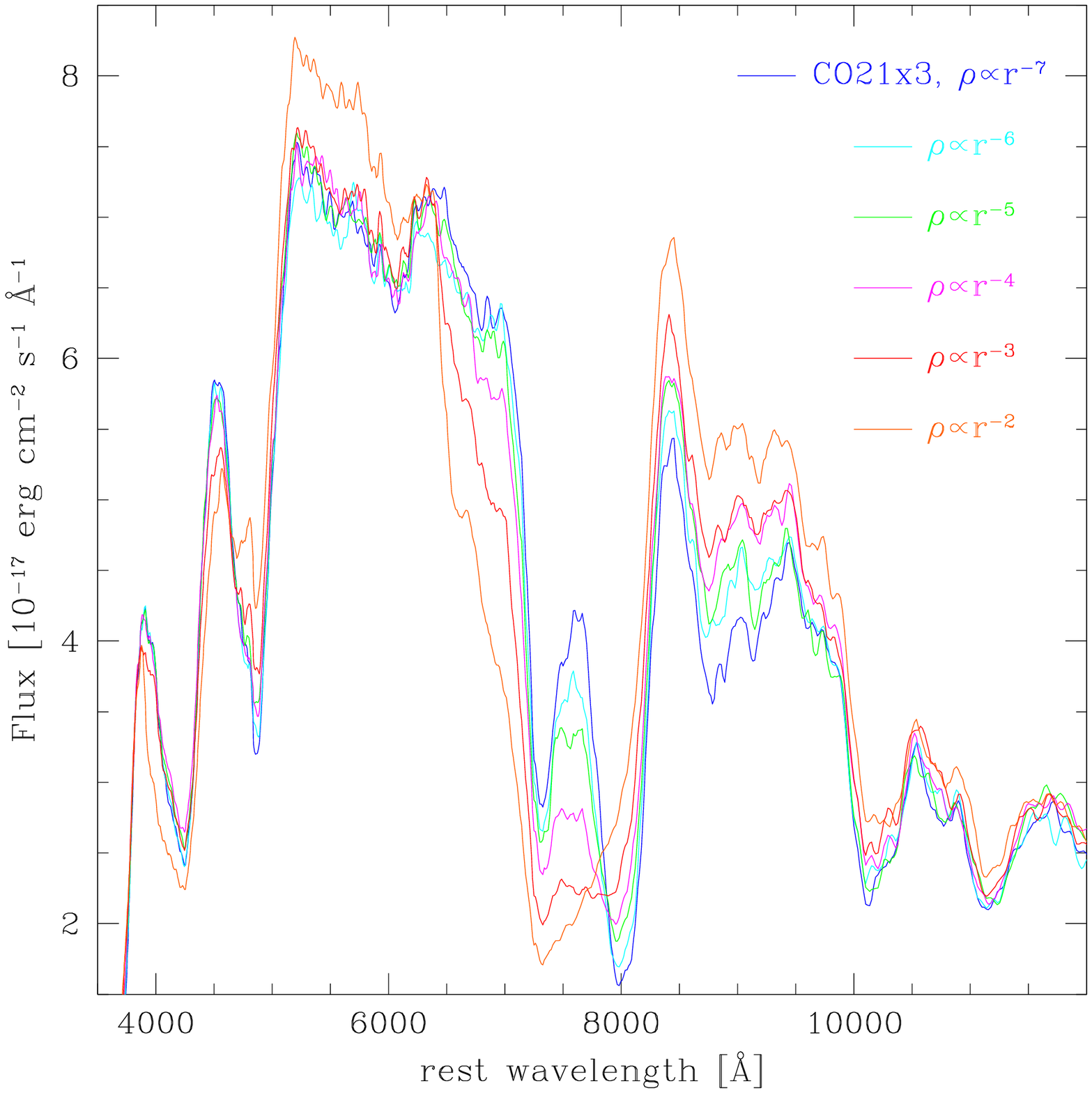}
  \caption{\label{fig:mockd7spec}Synthetic day 7 spectra obtained with the
  different input model density profiles used to test the earliest properties 
  of SN\,2004aw. }
\end{figure}

\begin{figure}
 \includegraphics[width=88mm]{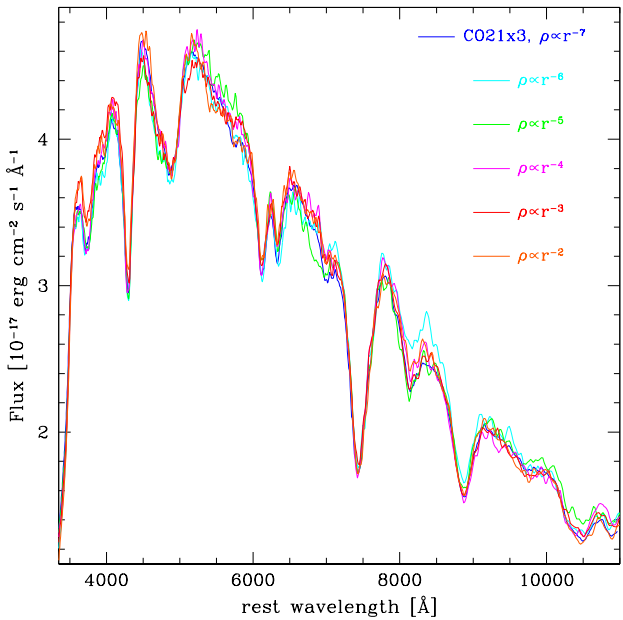}
  \caption{\label{fig:mockd15spec}Synthetic day 15 spectra obtained with the
  different input model density profiles used to test the earliest properties 
  of SN\,2004aw.}
\end{figure}

Figure \ref{fig:densityslope} shows the density profiles we have used. We
modified the density only above $v = 25,000$\,\kms\ in order not to affect the
models corresponding to epochs where observations exist showing that the lines
are not broad. The modified models have different density power-law indices
above that velocity, as marked in the figure. 
Figure \ref{fig:mockd7spec} shows the spectra at 7 days after explosion. Spectra
computed with increasingly flatter slopes show broader lines. These lines
eventually blend, reducing the number of observed features, as described in
\citet{Prentice2017a}. 
Figure \ref{fig:mockd15spec} shows the spectra at 15 days after explosion. At
this epoch all spectra are basically the same, showing that changing the outer 
density slope at high velocity has no impact at later times. 

The models we computed all have \Mej\,$\approx 4\,$\Msun, but vary in \KE\ from
$4 \times 10^{51}$\,erg for model CO21x3, which has the steepest outer density
slope, to $6.6 \times 10^{51}$\,erg for the model with outer density slope
power-law index $n=2$. The ratio \KE/\Mej\ ranges therefore from $\approx 1$ to
$\approx 1.6$, which spans some of the range of observed SN\,Ib/c properties,
although it does not reach the highest values. If \KE/\Mej\ was larger the
spectrum at 15 days would be affected. This sets the uncertainty on the \KE\
estimate for SN\,2004aw, and it is an uncertainty that should be applied to all
narrow-line SNe\,Ib/c with no early data. It also affects the detailed
classification of the SN, as the number of lines, which is 6 at maximum, can be
anything between 4 and 6 one week earlier \citet{Prentice2017a}.

Given these uncertainties, we can conservatively estimate that SN\,2004aw
ejected $4 \pm 1\,$\Msun\ of material with a kinetic energy \KE\,$\approx 4.5
\pm 1.5 \times 10^{51}$\,erg.

The value of $E_{\rm K}$ determined from modelling, although not as extreme as
the $\sim 10^{52}$ ergs of hypernovae, is probably too large for the canonical
``delayed neutrino-heating'' SN mechanism taking place in a proto-neutron star
\citep{jan07}. In fact, the ZAMS mass range of $\sim 23-30$\,M$_{\sun}$ spans
the putative upper boundary for neutron star formation at the end of core
collapse and a black hole may be preferred \citep{fry99}. However, doubts have
been cast on the sharpness of this separation \citep{ugliano12}. If the
progenitor core did collapse to a black hole, the SN explosion could have been
set off by a central engine comprising of the black hole and an accretion disk
as proposed in the collapsar model, providing that the progenitor star was
rotating rapidly \citep{mac99}. On the other hand, if the result of the collapse
was a neutron star, the explosion could have been aided by a magnetar, which may
have injected energy into the SN ejecta, contributed to the synthesis of \Nifs,
or both \citep{maz06a,mazzali14}. \citet{fryyou07} simulated the core collapse
of a 23 M$_{\sun}$ star and found a long delay to explosion, which may allow
time for large magnetic fields to develop in the proto-neutron star. 

\begin{figure}
  \centering
 \includegraphics[width=91mm]{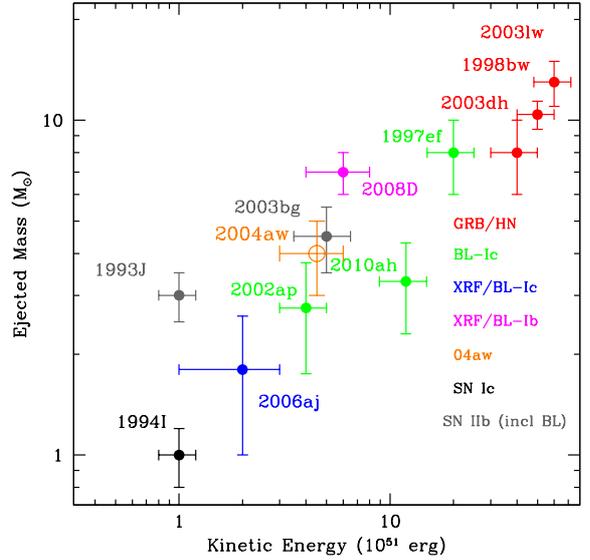}
  \caption{\label{fig:KEvMej}Kinetic energy vs. ejected mass in  
  SN\,2004aw (circles) and in other SNe\,Ic. }
\end{figure}

\begin{figure}
 \includegraphics[width=92mm]{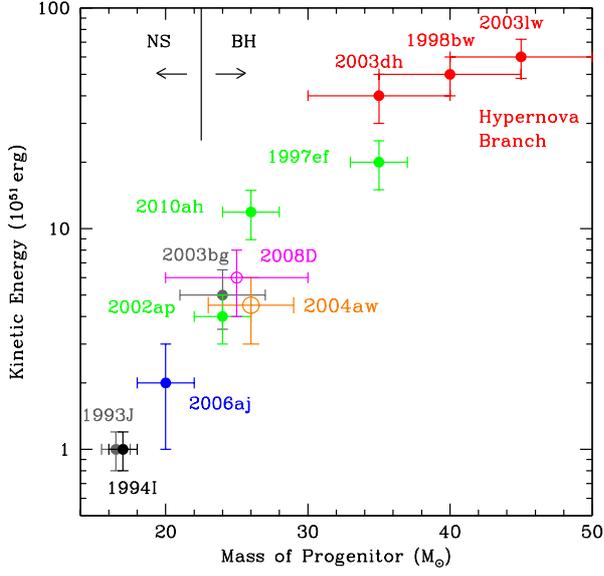}
  \caption{\label{fig:KEvM}Kinetic energy vs. inferred progenitor mass in  
  SN\,2004aw (circles) and in other SNe\,Ic. 
  Colour coding as in Fig. \ref{fig:KEvMej}.}
\end{figure}

Adding SN\,2004aw to the plots showing the main properties of SNe\,Ib/c that
have been modelled in detail (Figure \ref{fig:KEvMej}), we see that it confirms
the trend for increasing \KE\ with increasing ejected mass. This plot is
affected for some SNe by the presence of an outer He shell (SNe\,Ib) as well as
some H (SNe\,IIb). We can however use the mass of the CO core to reconstruct the
ZAMS mass of the progenitor. Such a plot (Figure \ref{fig:KEvM}) shows a more
linear relation between \KE\ and progenitor mass, although there seems to be
some spread in the value of \KE\ at progenitor masses between $\sim 20$ and
30\,\Msun. Among SNe in this mass range that have been studied, SN\,2004aw has
one of the smallest ratios of \KE\ per inferred progenitor mass, and it is in
fact the only one that does not show broad lines in its spectra.  

\begin{figure}
 \includegraphics[width=92mm]{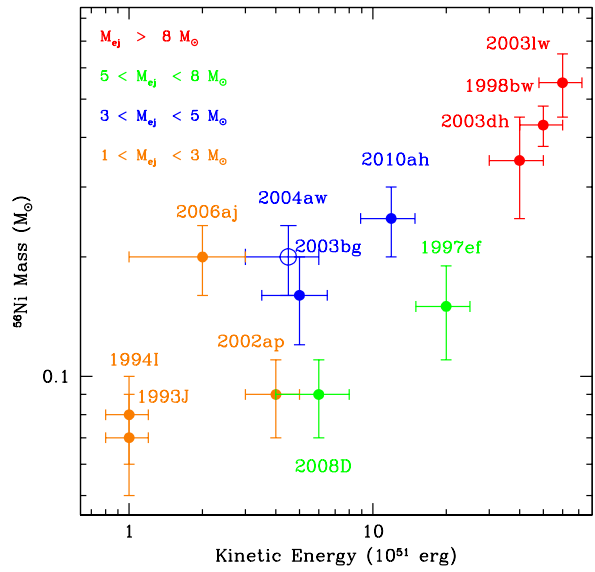}
  \caption{\label{fig:NivKE}\Nifs\ mass vs. kinetic energy in  
  SN\,2004aw (circles) and in other SNe\,Ic. }
\end{figure}

\begin{figure}
 \includegraphics[width=92mm]{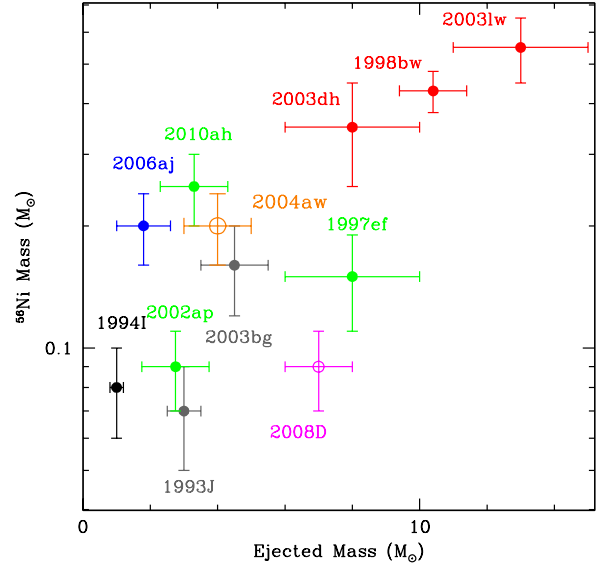}
  \caption{\label{fig:NivMej}\Nifs\ mass vs. ejected mass in  
  SN\,2004aw (circles) and in other SNe\,Ic. 
  Colour coding as in Fig. \ref{fig:KEvMej}.}
\end{figure}

\begin{figure}
 \includegraphics[width=92mm]{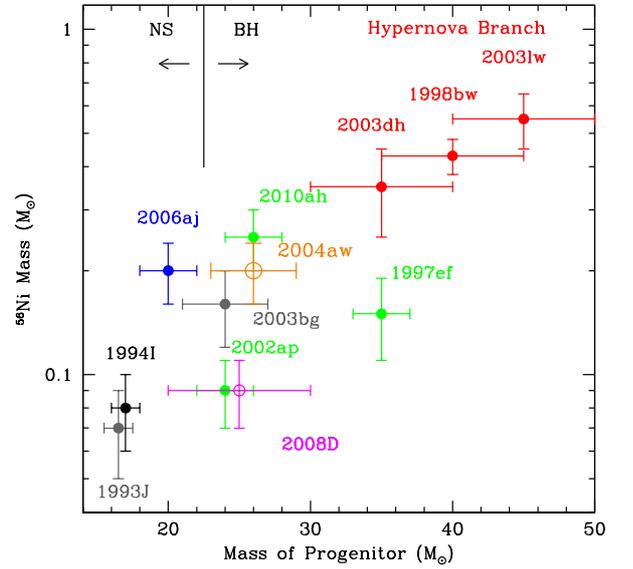}
  \caption{\label{fig:NivM}\Nifs\ mass vs. inferred progenitor mass in  
  SN\,2004aw (circles) and in other SNe\,Ic. 
  Colour coding as in Fig. \ref{fig:KEvMej}.}
\end{figure}

An amount of $\sim 0.20$ M$_{\sun}$ of $^{56}$Ni was ejected to power the light
curve of SN\,2004aw. Allowing for the uncertainties in the adopted distance
modulus and total reddening \citep{tau06} this value may vary from $\sim 0.15$
to 0.25 M$_{\sun}$. This value lies between the $\sim 0.3-0.6$ M$_{\sun}$
$^{56}$Ni of the three hypernovae connected with classical long GRBs (SNe
1998bw, 2003dh, and 2003lw; \citealt{nak01,den05,maz06b}) and the $\sim
0.07-0.1$ M$_{\sun}$ of the low-energy SN\,Ic 1994I \citep{iwa94} but
also of SN\,2002ap \citep{maz02,tom06}. On the other hand, it is comparable to
the value inferred for the Type Ic SN\,2006aj, which was accompanied by X-ray
Flash 060218 and for the non-GRB hypernova SN\,1997ef. SN\,2006aj has been
modelled to have had \KE\,$\approx 2\times 10^{51}$\,ergs and was suggested to
have been a magnetar-induced explosion of a star of $M_{\rm ZAMS}\sim 20$
M$_{\sun}$ \citep{maz06a}, while SN\,1997ef had $M_{\rm ZAMS}\sim 35$ M$_{\sun}$
and \KE\,$\approx 10^{52}$\,ergs \citep{iwa00,maz04}. The production of
explosively synthesized \Nifs\ appears broadly to increase with explosion energy
(Figure \ref{fig:NivKE}) as well as with ejected mass (Figure \ref{fig:NivMej}),
and the relation gets tighter if inferred progenitor mass is used, which
eliminates the influence of the outer stellar envelopes on the mass estimate
(Figure \ref{fig:NivM}). Still, in the range of values where SN\,2004aw is
located there is significant dispersion.  Several factors may be at play: in
SN\,2006aj magnetar energy may have contributed to increasing the \Nifs\
production. In the case of a massive progenitor such as that of SN\,1997ef,
fallback onto a black hole may have limited the amount of \Nifs\ that finally
was ejected. Other parameters, such as binarity, rotation, metallicity, and
asymmetry, may affect the outcome of the explosion. This highlights how little
we still know about SN\,Ib/c explosions. 

The properties of stripped-envelope core-collapse SNe seem particularly diverse
in the progenitor mass range $M_{\rm ZAMS}\sim 20-30$ M$_{\sun}$. While the
XRF-SN\,2006aj is strikingly similar to SN\,2002ap in light curves and spectra,
SN\,2002ap has no GRB association at all despite being one of the best-observed
SNe\,Ic. 

The Type Ib SN\,2008D, was captured by the Swift X-Ray Telescope at its very
initial flash of soft X-rays \citep{sod08} and was soon confirmed in the optical
\citep{den08}. With the nature of its initial flash under debate, spectrum and
light curve modelling suggests that it had a progenitor with $M_{\rm ZAMS}\sim
20 - 30$ M$_{\sun}$ \citep{maz08,tan09}.  Another peculiar SN\,Ib was 2005bf
($M_{\rm ZAMS}\sim 25-30$ M$_{\sun}$; \citealt{tom05}) whose composite light
curve reached a first peak compatible with moderate \Nifs\ production and then
rose for as long as $\sim 40$ days to reach a second peak, which would require
$\sim 0.3$ M$_{\sun}$ $^{56}$Ni. Late-time observations showed that the light
curve then traced back the predicted extension of a normal-luminosity SN,
suggesting that the second peak was due to the late injection of magnetar energy
\citep{maeda07}.  

Although our models are one-dimensional, asphericity would probably not change
our results much. Asphericity was observed in some Type Ic SNe with no GRB
connection, first using polarimetric measurements (e.g., \citealt{wan01,kaw02}),
and then also through revealing line profiles in nebular spectra (e.g,
\citealt{maz05,mae08}). The asphericity in SN\,2004aw is significant, but it
appears to be confined to the deepest layers of the SN ejecta, while in
SN\,1998bw it affected probably the entire SN ejecta.  As already mentioned,
what is suggested by the nebular spectra of SN\,2004aw is an explosion that was
fairly spherical in the outer layers, but significantly aspherical in the
deepest parts. It may heve been the result of a magnetar or a collapsar which
did not inject enough energy to modify the entire SN structure, but did leave an
imprint in the regions close to the site of collapse. This may be related to the
fact that SN\,2004aw had a smaller mass than any GRB/SN. Studies of further
examples of energetic SNe\,Ic are required in order to further our understanding
the presence and impact of an engine.

\section*{Acknowledgments} We gratefully acknowledge convertations with G.
Meynet during the KITP 2017 Programme on Massive stars. ST acknowledges support
by TRR33 ``The Dark Universe'' of the German Research Foundation.

\bsp
\label{lastpage}
\end{document}